\documentclass[fleqn,usenatbib]{mnras}

\usepackage{newtxtext,newtxmath}
\usepackage[T1]{fontenc}
\usepackage{ae,aecompl}

\usepackage{graphicx}	% Including figure files
\usepackage{amsmath}	% Advanced maths commands
\usepackage{amssymb}	% Extra maths symbols

\newcommand\blue{\color{blue}}
\newcommand\T{\rule{0pt}{2.6ex}}       % Top strut
\newcommand\B{\rule[-1.2ex]{0pt}{0pt}} % Bottom strut

\title[The galactic wind in IC~10 with LOFAR]{Exploring the making of a galactic wind in the star-bursting dwarf irregular galaxy IC~10 with LOFAR}

\author[V. Heesen et al.]{V.~Heesen,$^{1}$\thanks{E-mail: {\blue volker.heesen@hs.uni-hamburg.de}}
D.~A.~Rafferty,$^{1}$
A.~Horneffer,$^{2}$
R.~Beck,$^{2}$
A.~Basu,$^{2,3}$
J.~Westcott,$^{4}$
\newauthor 
L.~Hindson,$^{4}$
E.~Brinks,$^{4}$
K.~T.~Chy\.zy,$^{5}$
A.~M.~M.~Scaife,$^{6}$
M.~Br\"uggen,$^{1}$
\newauthor
G.~Heald,$^{7,8}$
A.~Fletcher,$^{9}$
C.~Horellou,$^{10}$
F.~S.~Tabatabaei,$^{11,12}$
R.~Paladino,$^{13}$
\newauthor
B.~Nikiel-Wroczy\'{n}ski,$^{4}$
M.~Hoeft,$^{14}$ and R.-J. Dettmar$^{15}$
\\
$^{1}$Universit\"at Hamburg, Hamburger Sternwarte, Gojenbergsweg 112, D-21029, Hamburg, Germany\\
$^{2}$Max-Planck-Institut f\"ur Radioastronomie, Auf dem H\"ugel 69, D-53121 Bonn, Germany\\
$^{3}$Fakult{\"a}t f{\"u}r Physik, Universit{\"a}t Bielefeld, Postfach 100131, D-33501 Bielefeld, Germany\\
$^{4}$School of Physics, Astronomy and Mathematics, University of
Hertfordshire, Hatfield AL10 9AB, UK\\
$^{5}$Astronomical Observatory, Jagiellonian University, ul. Orla 171, 30-244 Krak\'ow, Poland\\
$^{6}$Jodrell Bank Centre for Astrophysics, Alan Turing Building, Oxford Road, Manchester M13 9PL, UK\\
$^{7}$CSIRO Astronomy and Space Science, 26 Dick Perry Avenue, Kensington, WA 6151, Australia\\
$^{8}$University of Groningen, Kapteyn Astronomical Institute, NL-9700 AV Groningen, The Netherlands\\
$^{9}$School of Mathematics and Statistics, Newcastle University, Newcastle-upon-Tyne NE1 7RU, UK\\
$^{10}$Department of Space, Earth and Environment, Chalmers University of Technology, Onsala Space Observatory, SE-439 92 Onsala, Sweden\\
$^{11}$Instituto de Astrof\'isica de Canarias, V\'ia L\'actea S/N, E-38205 La Laguna, Spain\\
$^{12}$Departamento de Astrof\'isica, Universidad de La Laguna, E-38206 La Laguna, Spain\\
$^{13}$INAF/Istituto di Radioastronomia, via Gobetti 101, I-40129 Bologna, Italy\\
$^{14}$Th\"uringer Landessternwarte (TLS), Sternwarte 5, D-07778 Tautenburg, Germany\\
$^{15}$Astronomisches Institut der Ruhr-Universit\"at Bochum, D-44780 Bochum, Germany
}

\date{Accepted 2018 February 01. Received 2018 January 26; in original form 2017 November 07.}

\pubyear{2018}

\begin{document}
\label{firstpage}
\pagerange{\pageref{firstpage}--\pageref{lastpage}}
\maketitle

\begin{abstract}
Low-mass galaxies are subject to strong galactic outflows, in which cosmic rays may play an important role; they can be best traced with low-frequency radio continuum observations, which are less affected by spectral ageing. We present a study of the nearby star burst dwarf irregular galaxy IC~10 using observations at 140~MHz with the LOw-Frequency ARray (LOFAR), at 1580~MHz with the Very Large Array (VLA) and at 6200~MHz with the VLA and the 100-m Effelsberg telescope. We find that IC~10 has a low-frequency radio halo, which manifests itself as a second component (thick disc) in the minor axis profiles of the non-thermal radio continuum emission at 140 and 1580~MHz. These profiles are then fitted with 1D cosmic-ray transport models for pure diffusion and advection. We find that a diffusion model fits best, with a diffusion coefficient of $D=(0.4$--$0.8) \times 10^{26}(E/{\rm GeV})^{0.5}~{\rm cm^2\,s^{-1}}$, which is at least an order of magnitude smaller than estimates both from anisotropic diffusion and the diffusion length. In contrast, advection models, which cannot be ruled out due to the mild inclination, while providing poorer fits, result in advection speeds close to the escape velocity of $\approx$$50~\rm km\,s^{-1}$, as expected for a cosmic-ray driven wind. Our favoured model with an accelerating wind provides a self-consistent solution, where the magnetic field is in energy equipartition with both the warm neutral and warm ionized medium with an important contribution from cosmic rays. Consequently, cosmic rays can play a vital role for the launching of galactic winds in the disc--halo interface. 
\end{abstract}

%Our estimated advective time-scale of 25--50~Myr is in good agreement with the 30-Myr age of the oldest stellar cluster, suggesting that the wind has been launched during the current starburst.


%%%%%%%%%%%%%%%%%%%%%%%%%%%%%%%%%%%%%%%%%%%%%%%%%%

%%%%%%%%%%%%%%%%% BODY OF PAPER %%%%%%%%%%%%%%%%%%

\section{Introduction}

Galactic winds are an important ingredient in galaxy evolution, removing angular momentum and energy over a galaxy's lifetime \citep[e.g.][]{veilleux_05a}. Such winds are thought to be prevailing in dwarf irregular galaxies, which have burst-like star-formation histories and weak gravitational potentials \citep{tremonti_04a}. Because dwarf galaxies are the building blocks of galaxy formation in the early Universe and are the most abundant type of galaxies, it has been suggested that galactic winds in them can explain the metal enrichment of the intergalactic medium at high redshifts \citep{recchi_13a}. Cosmic rays have long been suspected to play a part in the launching of galactic winds, their cooling  being much slower than that of the hot, supernovae-heated gas \citep{breitschwerdt_91a}. This idea has recently been addressed again in several theoretical papers
\citep[e.g.][]{Pakmor_16} with the conclusion that cosmic rays could play an important role in launching galactic winds. On the observational side this can be explored by studying the synchrotron emission of cosmic-ray electrons (CREs) spiralling around the magnetic field lines -- this emission is the non-thermal component of the radio continuum emission from galaxies.

Aside from this aspect of studying the importance of cosmic rays for the physics of the ISM, radio continuum emission in galaxies is an important tracer for the star-formation rate (SFR), un-obscured by dust and observable with ground-based telescopes \citep[][]{heesen_14a,tabatabaei_17a}. Radio interferometers have the necessary angular resolution in order to study nearby galaxies at 100-pc scales that allows us to explore the physical processes behind the radio--SFR relation. At GHz- and even MHz- frequencies, where the non-thermal emission dominates over the thermal (free--free) emission, a galaxy may be either (i) an \emph{electron calorimeter}, i.e.\ the cosmic-ray electrons (CREs) are confined within the galaxy and radiate all their energy away \citep{voelk_89a}; or (ii) there may be \emph{energy equipartition} between the cosmic rays and the magnetic field as well as a relation between the magnetic field strength ($B$) and the SFR \citep{niklas_97a}. Of course, intermediate states may also exist. Present-day observations of spiral galaxies seem to favour model (ii) and galactic winds can be a mechanism to remove CREs and attain energy equipartition. Furthermore, a $B$--SFR relation can be realized in an elegant way, by assuming a constant velocity dispersion in the interstellar medium (ISM), which is a good approximation, and a Kennicutt--Schmidt relation between the SFR and the gas density ($\rho$). A $B$--$\rho$, and thus a $B$--SFR, relation is then a consequence if the magnetic energy density is in equipartition with the kinetic energy density of the ISM \citep{heesen_14a}. 

In this paper, we present a low-frequency radio continuum study of the nearby star burst dwarf irregular galaxy IC~10 to shed new light on the interplay between magnetic fields, cosmic rays and star formation. In particular, we would like to explore whether a radio halo exist and, if so, whether it can be explained with a cosmic ray-driven wind such as found in late-type spiral edge-on galaxies \citep{heesen_18b}. Owing to its proximity ($D=0.7$~Mpc), this galaxy is particularly well suited for such an investigation since we can study the ISM at sub-kpc scales ($\rm 1~arcsec\approx 3.39$~pc). Low frequencies are advantageous for such work. First, dwarf irregular galaxies have a high thermal contribution to their radio continuum emission \citep[typically 30 per cent at $1.4$ GHz rather than 10 per cent as in spiral galaxies;][]{tabatabaei_17a}. Hence, going to $\approx$140~MHz is expected to reduce the thermal fraction to a 10- or 20-per-cent level; this is because the thermal radio continuum emission depends only weakly on frequency ($\propto \nu^{-0.1}$), whereas the non-thermal part increases at low frequencies ($\propto \nu^{-0.6}$, typically). Second, low frequencies allow us to explore the synchrotron emission of CREs far away from star-formations sites that cannot be detected at GHz-frequencies due to spectral ageing. Such a study has now become possible with the advent of the LOw-Frequency ARray \citep[LOFAR;][]{vanHaarlem_13a}, which combines high angular resolution ($\approx$10~arcsec) with a sensitivity  that is approximately matched with other state-of-the-art interferometers operating at GHz-frequencies such as the Karl G.\ Jansky Very Large Array (VLA).

IC~10 is considered a dwarf irregular galaxy \citep[dynamical mass of $1.6\times 10^8~\rm M_{\sun}$;][]{oh_15a} currently undergoing a star burst, owing to its high number of Wolf--Rayet stars \citep{massey_95a}. It has a large supply of atomic hydrogen, detected as \ion{H}{i} line emission in the radio, suggesting that the galaxy has accreted a large supply of hydrogen in recent times, which is able to sustain its current high star-formation surface density $\Sigma_{\rm SFR}\approx 10^{-1}~\rm M_{\sun}\,yr^{-1}\,kpc^{-2}$ \citep{hunter_12a}. IC~10 has been studied extensively in the radio continuum as well. \citet{chyzy_16a} showed that the galaxy has ordered magnetic fields, which are arranged in a similar pattern (X-shaped) as has been found in nearby edge-on spiral galaxies. They suggest that the galaxy has a spherical outflow as found in MHD simulations of dwarf galaxies. IC~10 has several massive \ion{H}{ii} regions, of which the south-eastern star-formation complex hosts a 150-pc non-thermal superbubble, likely the result of a few supernovae or one massive hypernova in the last few Myr \citep{heesen_15a}. A recent low-frequency study of \citet{basu_17a} concentrated on the spatially resolved radio--SFR relation; they showed that the relation holds down to 50-pc spatial resolution in the disc close to the star-formation sites. \citet{westcott_17a} used high-resolution radio continuum observations to investigate the nature of the compact sources in IC~10; they were  shown to be either background active galactic nuclei (AGNs) or internal \ion{H}{ii} regions.

This paper is organized as follows: In Section~\ref{sec:observations}, we outline our observations and data reduction of our new LOFAR observations. Section~\ref{sec:results} presents our results together with those from archival VLA and Effelsberg data. In Section~\ref{sec:cr_transport}, we model the radio continuum data with 1D cosmic-ray transport in order to measure diffusion coefficients and advection speeds. These models, which are either for pure diffusion along magnetic field lines or for pure advection in a galactic wind, allow us to separate the cosmic ray and magnetic energy densities without the assumption of (local) energy equipartition. In Section~\ref{sec:discussion}, we then study the relation between cosmic rays, magnetic fields and the other constituents to the ISM such as the warm neutral and warm ionized medium. We summarize and present our conclusions in Section~\ref{sec:conclusions}.

\section{Observations}
\label{sec:observations}
Observations with the LOFAR high-band antenna (HBA) system were taken on 2013 August 25 and 26 in the HBA\verb|_|DUAL\verb|_|INNER configuration as part of the observing programme LC0\verb|_|43. The observations were taken in interleaved mode, with alternating scans of the calibrator 3C~48 (2~min) and the target (12~min). The data were reduced with the novel `facet calibration' technique, which mitigates the direction-dependent effects of the ionosphere and beam response that impact low-frequency radio continuum observations with aperture arrays, so that images close to the thermal noise level can be obtained \citep{van_weeren_16a,williams_16a}. First, the $(u,v)$ data are calibrated with direction-independent methods using the {\small PREFACTOR} pipeline.\footnote{\href{https://github.com/lofar-astron/prefactor}{https://github.com/lofar-astron/prefactor}} 
This pipeline first calibrates 3C~48 using the \citet{scaife_12a} flux densities, assuming a 
point-like source. From the resulting gain solutions the instrumental components are extracted:
the station gain amplitudes and the phase variations due to the drift of the clocks of the LOFAR 
stations. The latter is separated from the variations due to the changing total electron 
content (TEC) of the ionosphere with the clock--TEC separation.
These instrumental calibration solutions are applied to the target data, which are then 
averaged to 10-s time resolution and 2-channels-per-sub-band frequency resolution (channel 
width of $97.656~\rm kHz$). The data are calibrated in phase only using the Global Sky Model~\citep[GSM;][]{scheers_11a}, which is a compilation of sources from the VLA Low-frequency Sky Survey Redux \citep[VLSSr;][]{lane_14a}, the Westerbork Northern Sky Survey \citep[WENSS;][]{rengelink_97a} and the NRAO VLA Sky Survey \citep[NVSS;][]{condon_98a}.  With the direction-independent calibration applied, the $(u,v)$ 
data are inverted and deconvolved with a wide-field {\small CLEAN} algorithm. As final step of {\small PREFACTOR}, the {\small CLEAN} components of all the sources within the 
$\approx$$8\degr$ field of view (FOV) are subtracted from the $(u,v)$ data.
\begin{figure*}
	\includegraphics[width=\hsize]{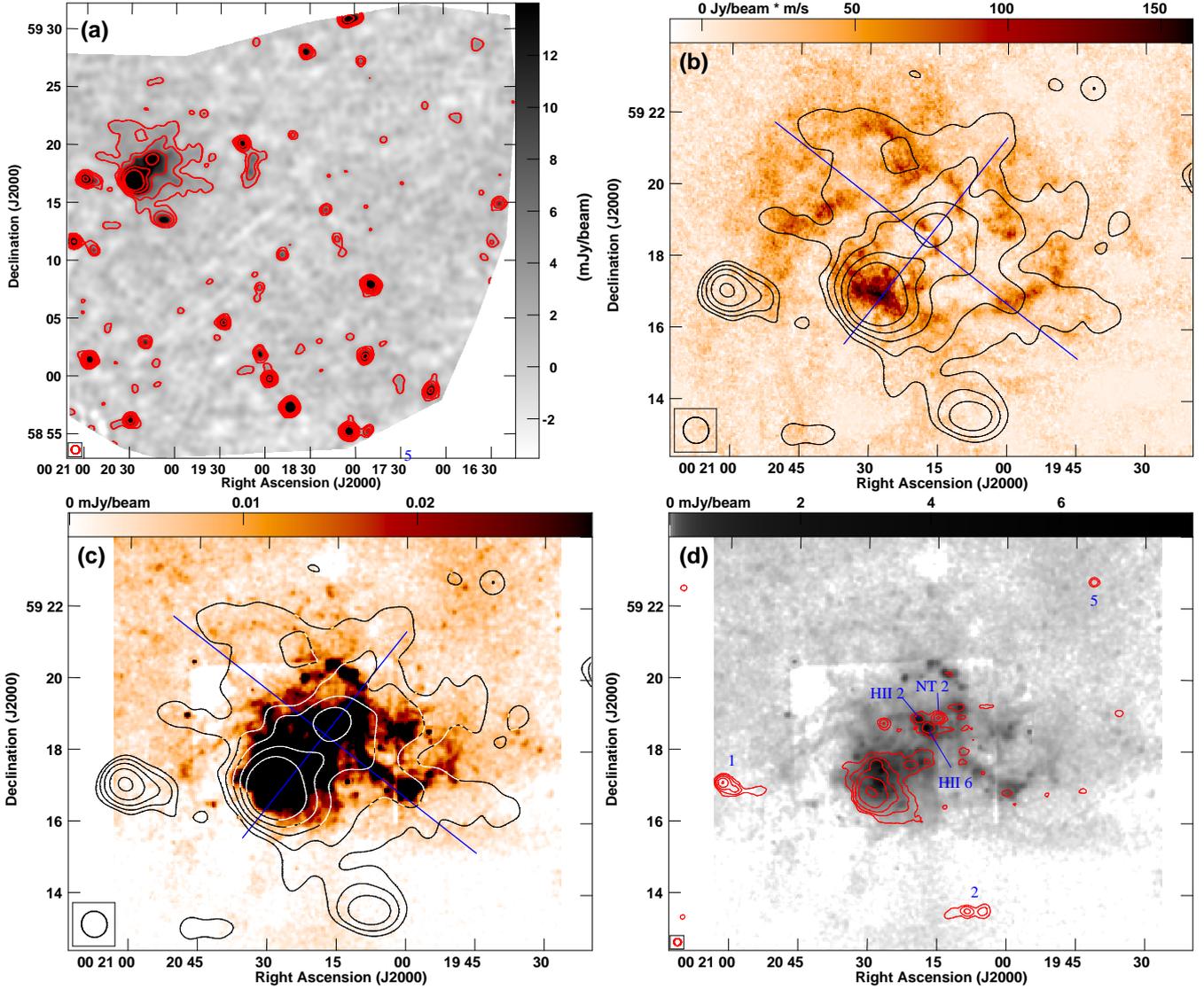}
    \caption{(a) Full target facet, where the LOFAR 140-MHz radio continuum emission at $44.0\times 42.3~\rm arcsec^2$ resolution is shown as contours, overlaid on the same emission as grey-scale. Contours are at $2^n\times 3\sigma$, with $n=0,1,2\ldots 6$ and $\sigma=700~\umu\rm Jy\,beam^{-1}$. (b) Zoom in on IC~10, showing an area of approximately $2.9\times 2.3~\rm kpc^2$, where 140-MHz contours are overlaid on an integrated \ion{H}{i} emission map from LITTLE THINGS at $5.9\times 5.5~\rm arcsec^2$ resolution. Contour levels as in panel (a). (c) Same area as in (b), where 140-MHz contours are overlaid on the thermal 140-MHz radio continuum emission at $6\times 6~\rm arcsec^2$ resolution. The thermal radio continuum was estimated from a combination of H\,$\alpha$, \emph{Spitzer} 24-$\umu$m and VLA 32-GHz emission. Contour levels as in panel (a). (d) Same area as in (b), where LOFAR 140-MHz contours at a resolution of $12\times 12~\rm arcsec^2$ are overlaid on the 140-MHz thermal radio continuum emission as in panel (c). Contour levels are at $2^n\times 4\sigma$, with $n=0,1,2,3,4$ and $\sigma=300~\umu\rm Jy\,beam^{-1}$. In panels (b) and (c), blue lines show the major and minor axes of the radio continuum emission. Note that the shorter line is coincident with the optical major axis ($\rm PA = -38\degr$). In all panels, the size of the synthesized beam of the LOFAR map is shown in the bottom-left corner.}
    \label{fig:maps}
\end{figure*}

The residual, direction-independent calibrated $(u,v)$ data with all sources subtracted, together with the subtracted model and the solutions of the phase-only calibration, are then the input for the direction-dependent facet calibration, for which we used the {\small FACTOR} pipeline (Rafferty et al.\ 2018, in prep.).\footnote{\href{https://github.com/lofar-astron/factor}{https://github.com/lofar-astron/factor}} The FOV was divided into 55 facets around calibrator regions with integrated 177-MHz flux densities (of the full facet) between $0.3$ and $12$~Jy.  Of those, the 7 brightest facets were processed one at a time, with the faintest facet having a flux density of $2.1$~Jy. The facet calibration technique allows us to track and correct for the direction-dependent effects of the Earth's ionosphere (effectively the `seeing' at long radio-wavelengths) and the station beam response by first self-calibrating on the calibrator region of a facet and then using the solutions to update the model for the full facet, which in turn is used to update the residual $(u,v)$ data. 
In the first step of the calibration, fast, 10-s phase solutions are determined in small chunks of $\approx$2~MHz bandwidth to correct for the positional change and distortion of sources. In the second step, slow, tens-of-minutes amplitude solutions are used to track the variation of the apparent flux density of a source. The target facet was corrected using the solution of a nearby, $0\fdg 9$ away, facet at $\rm RA~00^h15^m50\fs  76$, $\rm Dec.~58\degr36\arcmin23\farcs 47$ (J$2000.0$) with a 177-MHz flux density of $5.4$~Jy.

For the final data set we used observations of 23 core stations, each acting as 2 stations, and 13 remote stations, $5.4$~h of on-source time with $49.2$~MHz bandwidth. The direction-dependent calibrated $(u,v)$ data were taken into the Common Astronomy Software Applications \citep[{\small CASA};][]{mcmullin_07a} and inverted and deconvolved with the MS--MFS {\small CLEAN} algorithm \citep{rau_11a}. We fitted for the frequency dependence of the skymodel ($\rm nterms=2$) and used angular scales of up to 600~arcsec diameter. The $(u,v)$-range was restricted to $0.14\rightarrow 22.0$~k$\lambda$, in order to match the $L$-band VLA observations in B- and C-arrays (see below); we used an outer taper of $\approx$$40~\rm arcsec$ with Briggs weighting, setting $\rm robust=0$. This resulted in a 140-MHz image with $44.0\times 42.3~\rm arcsec^2$ resolution ($\rm PA=17\fdg 4$) with a rms noise level of $700~\umu\rm Jy\,beam^{-1}$; this image of the target facet is presented in Fig.~\ref{fig:maps}(a).\footnote{Angular resolutions in this paper are referred to as the full width at half maximum (FWHM).} This is the map we use in what follows unless mentioned otherwise. We also created a high-resolution map using all $(u,v)$ data and no taper with a resolution of $12.0\times 12.0~\rm arcsec^{2}$ and a rms noise level of $300~\umu\rm Jy\,beam^{-1}$.

In addition, we created a 1580-MHz map from VLA observations in B- and C-arrays, already presented in \citet{heesen_15a}. We used the same techniques as for the LOFAR 140-MHz image; the resulting map has a resolution of $43.3\times 41.4~\rm arcsec^2$ ($\rm PA = 27\fdg 8$) with a rms noise level of $70~\umu\rm Jy\,beam^{-1}$. We also used a 6200-MHz map from VLA D-array observations, combined with observations from the 100-m Effelsberg telescope \citep{basu_17a}. This map has a resolution of $9.4\times 7.3~\rm arcsec^2$ ($\rm PA=-54\fdg 6$) and a rms noise level of $15~\umu\rm Jy\,beam^{-1}$.

\section{Results}
\label{sec:results} % used for referring to this section from elsewhere

In Fig.~\ref{fig:maps}(b), we show an overlay of our 140-MHz map as contours on a map of the integrated \ion{H}{i} emission-line intensity from the LITTLE THINGS survey \citep[Local Irregulars That Trace Luminosity Extremes, The \ion{H}{i} Nearby Galaxy Survey;][]{hunter_12a} as colour-scale. The same contours are overlaid in Fig.~\ref{fig:maps}(c) on the map of the thermal radio continuum emission, also at 140~MHz, which was constructed from H\,$\alpha$ emission line maps \citep{gil_de_paz_03a, hunter_04a} and corrected for foreground and internal absorption (see below). Figure~\ref{fig:maps}(d) shows this map with a logarithmic scaling, where we overlaid 140-MHz contours at 12~arcsec resolution. As can be seen, the low-frequency radio continuum emission does not correspond morphologically to either the warm neutral or the warm ionized medium. The radio continuum emission extends significantly into the halo with an extent of the $3\sigma$-contour line of  $10.7$~arcmin ($2.2$~kpc) along the minor axis and  $7.3$~arcmin ($1.5$~kpc) along the major axis \citep[$\rm PA=-38\degr$;][]{hunter_12a}, as shown by the blue lines in Figs~\ref{fig:maps}(b) and (c). Assuming an inclination angle of $i=47\degr$ \citep{oh_15a}, we would expect a minor axis extension of only $\approx$5~arcmin ($\approx$$1.0$~kpc). 

\begin{figure}
	\includegraphics[width=\hsize]{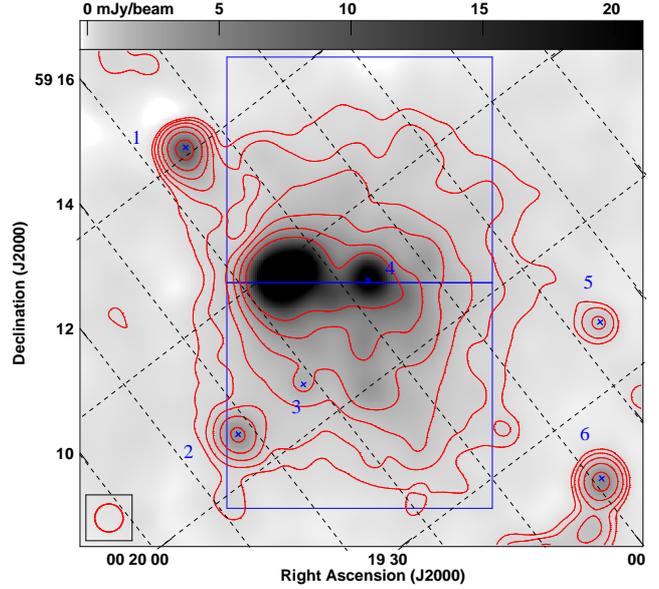}
    \caption{Non-thermal radio continuum emission at 1580~MHz with an angular resolution of $44.0\times 44.0~\rm arcsec^2$. Contours are $2^n\times 3\sigma$, with $n=0,1,2\ldots 5$ and $\sigma=70~\umu\rm Jy\,beam^{-1}$. The map has been rotated, so that the major axis of the radio continuum emission is horizontal. The minor axis strip used for the averaging along with the masked sources are shown as well. The size of the synthesized beam is shown in the bottom-left corner, which corresponds to a spatial resolution of $\approx$150~pc.}
    \label{fig:20cm}
\end{figure}

We do not find any indication for the spherical `synchrotron envelope', which \citet{chyzy_16a} detected using 1430-MHz VLA observations. This can be explained by the flat radio spectral index in the halo of IC~10. The integrated 140-MHz flux density of IC~10 is $756\pm 29$~mJy, where we excluded sources 1 and 2 (Fig.~\ref{fig:maps}d), which are unrelated background sources; source 1 has the morphology of a head-tail radio galaxy and source 2 that of a Fanaroff--Riley type II radio galaxy \citep[FR;][]{fanaroff_74a}. The integrated 1580-MHz flux density is $306\pm 10$~mJy, resulting in a 140--1580~MHz spectral index of $-0.38\pm 0.02$.\footnote{Source 3 (see Fig.~\ref{fig:20cm}), likely a background AGN \citep{westcott_17a}, is included in this measurement since it is marginally detected only at 12 arcsec resolution ($3.5\sigma$) but cannot be separated from the background emission at 44 arcsec resolution.} Thus, the increased noise level in our LOFAR observations, compared with the VLA observations of \citet{chyzy_16a}, is not compensated by a steep spectrum and our data are less sensitive by comparison.

We subtract the contribution from the thermal radio continuum emission with a combination of H\,$\alpha$, \emph{Spitzer} 24-$\rm\umu m$ mid-infrared and 32-GHz radio continuum emission \citep{heesen_15a, basu_17a}. The foreground absorption was corrected with $A_V=2.59\times E(B-V)$ using $E(B-V)=0.75~\rm mag$ \citep{burstein_84a}. The correction for internal absorption was done with the \emph{Spitzer} 24-$\rm\umu m$ mid-infrared data \citep{kennicutt_09a} using the flux densities of \citet{bendo_12a}. This extinction corrected H\,$\alpha$ map was then combined with a 32-GHz VLA map of the south-eastern H\,{\sc ii} region \citep[see][for details]{heesen_15a}. This results in an integrated thermal flux density of $161\pm 5~\rm mJy$ at 140~MHz. Hence, the non-thermal integrated radio spectral index between 140 and 1580~MHz is $-0.50\pm 0.03$, which is consistent with the radio spectral index of $\approx$$-0.5$, expected for freshly injected CREs. The radio spectral index is also marginally consistent with the power-law fit between 320 and 24500~MHz \citep[$0.55\pm 0.04$;][]{basu_17a} and the injection radio spectral index in the non-thermal superbubble \citep[$0.55\pm 0.05$;][]{heesen_15a}.

There are several reasons why the non-thermal radio continuum spectrum may deviate from a power law,  showing a flattening towards low frequencies. First, below $\approx$1~GHz, some of the compact radio continuum sources show a spectral turn-over (increasing flux density with frequency) due to thermal (free--free) absorption \citep{basu_17a}. This is in particular the case for source 4 (Fig.~\ref{fig:20cm}), which is unresolved at  $44.0\times 42.3~\rm arcsec^2$ resolution but can be resolved into three components at $12.0\times 12.0~\rm arcsec^2$ resolution, two of which are \ion{H}{ii} regions (sources \ion{H}{ii} 2 and 6; Fig.~\ref{fig:maps}d) and one is an unrelated non-thermal background AGN \citep[source NT 2;][]{westcott_17a}. Source \ion{H}{ii} 2 shows a marginal spectral downturn at 320~MHz \citep{basu_17a}, which does not manifest itself at 140~MHz. The downturn of source H\,{\sc ii} 6, however, can be confirmed; it has a 140-MHz flux density of only $\approx$4~mJy, whereas the expected flux density is $\approx$14~mJy if the source would be following a power law. Still, the difference in spectral index is only $\Delta \alpha = +0.01$, which is not enough to explain the spectral flattening. In the diffuse ISM, thermal absorption is not expected to play a role. A second reason could be the escape of low-energy CREs in a galactic wind, which should be visible as a radio halo.

\begin{figure}
	\includegraphics[width=\hsize]{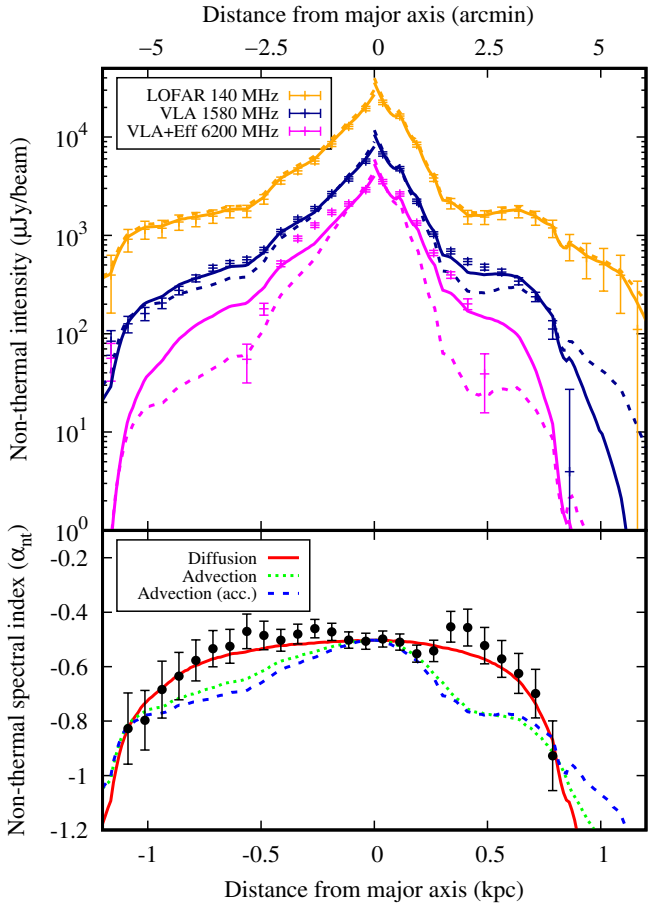}
    \caption{Minor axis profiles with distances from the major axis as seen on the sky. In the \emph{top panel}, data points show the non-thermal radio continuum intensities at 140, 1580 and 6200~MHz. Solid lines show the best-fitting diffusion model (Model~I, $\umu=0.5$) and dashed lines our favoured model, which is advection with an accelerating wind (Model~III). Note that the 140-MHz intensities are almost identical for both models. In the \emph{bottom panel}, data points show the non-thermal radio spectral index between 140 and 1580~MHz. The various lines show the best-fitting models, where the solid red line shows the diffusion model (Model~I, $\umu=0.5$), the short-dashed green line the advection model with constant wind speed (Model~II) and the long-dashed blue line the advection model with an accelerating wind (Model~III). All models assume $\gamma_{\rm inj}=2.0$ and the north-eastern halo is on the right-hand side ($z>0$~kpc) and the south-western halo on the left-hand side ($z<0$~kpc). Error bars show $\pm 1\sigma$ uncertainties as calculated in Section~\ref{sec:results}.}
    \label{fig:alpha}
\end{figure}

In order to investigate the existence of extra-planar emission and hence that of a radio halo further, we created minor axis profiles of the non-thermal radio continuum emission. We convolve the non-thermal radio continuum maps at 140, 1580 and 6200~MHz to $44.0\times 44.0~\rm arcsec^2$ resolution and rotate them, so that the radio major axis lies horizontal. As an example, we show in Fig.~\ref{fig:20cm} the 1580-MHz map. Again, as with the 140-MHz map, the emission extends prominently along the minor axis. The contours are box-shaped rather than of elliptical shape, so that emission away from the major axis could be extra-planar emission that is located above the star-formation sites in the disc. This is the sign for the existence of a radio halo, where the emission extends vertically away from the star-forming disc but not radially \citep{dahlem_06a}. We have averaged the radio continuum emission within a strip along the minor axis, centred on $\rm RA~00^h20^m17\fs 732$ $\rm Dec.~59\degr 18\arcmin 27\farcs 24$, with a width of $6.9$~arcmin ($1.4$~kpc) and a height a height of $11.8$~arcmin ($2.4$~kpc), spacing data points by 22~arcsec ($\approx$$75$~pc, equivalent to FWHM/2) and masking unrelated background sources and those with spectral turn-overs (sources 1--4). In addition, we masked source 5, which is unresolved, and source 6, which has the morphology of a FR~II radio galaxy; both have non-thermal spectral indices ($\approx$$-0.6$) and are likely unrelated to IC~10. Error estimates of the intensities were calculated as $\delta I_{\nu}^2=(\epsilon_{I} I_{\nu})^2+\sigma_{\rm b}^2$, where $\epsilon_{I}=0.05$ is the calibration uncertainty and $\sigma_{\rm b}=\sigma/\sqrt{N}$ is the intensity baselevel uncertainty with $N$ the number of beams within the integration region. The non-thermal radio spectral index error is then:
\begin{equation}
	\delta\alpha_{\rm nt} = \frac{1}{\left | \ln\left ( \frac{\nu_1}{\nu_2} \right ) \right |} \sqrt{\left ( \frac{\delta I_1}{I_1} \right )^2 + \left ( \frac{\delta I_2}{I_2}\right )^2},
\end{equation}
where $I_1$ and $I_2$ are the non-thermal radio continuum intensities at frequencies $\nu_1$ and $\nu_2$, respectively.

The resulting minor axis profiles of the non-thermal radio continuum emission at 140, 1580 and 6200~MHz and of the non-thermal radio spectral index between 140 and 1580 MHz are presented in Fig.~\ref{fig:alpha}. As can be seen, the 140- and 1580-MHz profiles show a good correspondence everywhere. We detect a break in the intensity profiles at a major axis distance of $\approx$$0.5$~kpc, where the decrease of the intensity with increasing distance significantly flattens. We interpret this as the transition from the projected galactic disc to the halo emission, where the thick disc corresponds to the halo emission. The distance of $0.5$~kpc from the major axis corresponds to the projected length of the semi-major axis, $\cos(i)\times 1.5~{\rm kpc}/2$, which is the distance from the major axis up to which emission from the galactic disc still contributes. The 6200-MHz profile can be only traced out to $0.5$~kpc distance from the major axis, at larger distances the emission decreases rapidly. Hence, we do not detect a radio halo at 6200~MHz. At all frequencies, the profiles show the expected behaviour for spectral ageing.

\begin{figure}
	\includegraphics[width=\hsize]{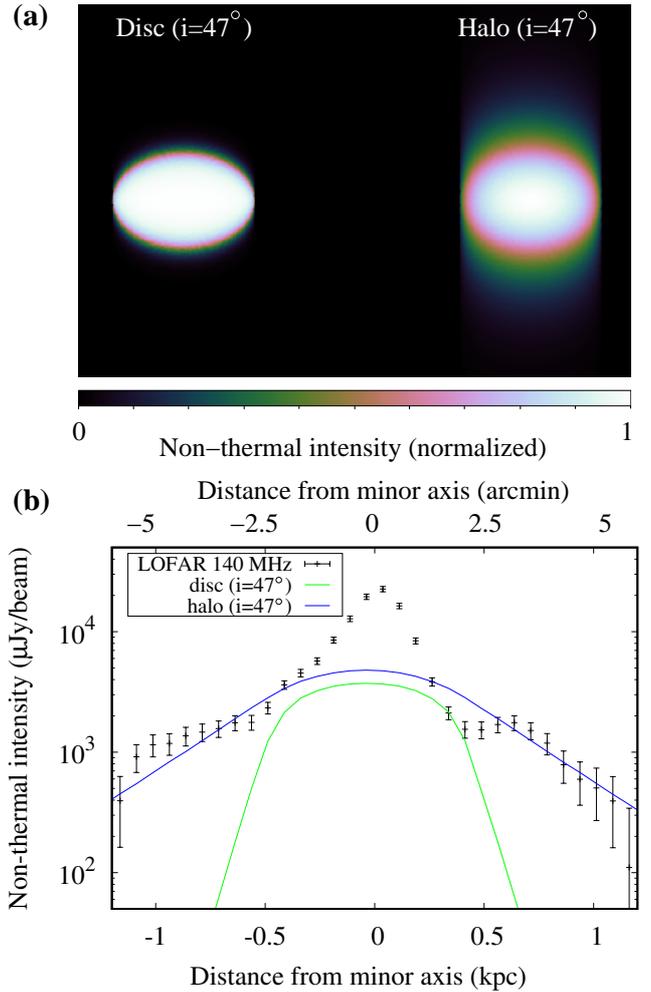}
    \caption{(a) synthetic non-thermal intensity maps from a 3D toy model of an inclined cylinder of synchrotron emissivity with a vertical intrinsic scale height of either $0.1$~kpc (disc model) or $0.5$~kpc (halo model). (b) comparison of the synthetic minor axis profiles of the intensity (arbitrary scaling) with the 140-MHz non-thermal intensity profile (as in Fig.~\ref{fig:alpha}). Our toy model does not trace the excess emission at distances $<$$0.5~\rm kpc$, which is due to the non-thermal superbubble that is not included in our model.}
    \label{fig:model}
\end{figure}

The existence of a radio halo can be further corroborated by a 3D toy model, which is then converted into synthetic intensity maps. We created a `disc' model that has a vertical intrinsic intensity scale height of $0.1$~kpc and a `halo' model with a vertical intrinsic intensity scale height of $0.5$~kpc. The synchrotron emissivity is constant for galactocentric radii $\leq$$0.7$~kpc and vanishes outside of this cylinder with a diameter of $1.4$~kpc. These cylindrical models were tilted away from edge-on, corresponding to the inclination angle of $i=47\degr$, and the emissivity was integrated along the line of sight. The resulting synthetic intensity distribution and minor axis intensity profile are shown in Fig.~\ref{fig:model}. Clearly, the disc model cannot explain the observed intensity profile at distances $>$$0.5$~kpc from the major axis, whereas the halo model can. Thus, the existence of a halo is required to account for the observations.

\section{Cosmic-ray transport}
\label{sec:cr_transport}
We now  use the low-frequency radio spectral index information in order to study the transport of CREs away from the star-formation sites, which are located near the galactic midplane. We have used the software {\small SPINNAKER} (Spectral Index Numerical Analysis of K(c)osmic-ray Electron Radio-emission), which calculates vertical synchrotron emission profiles for stationary 1D cosmic-ray transport models for either pure advection or diffusion \citep{heesen_16a}.\footnote{\href{https://github.com/vheesen/Spinnaker}{https://github.com/vheesen/Spinnaker}} We assume that the CREs are injected near the major axis and are transported vertically away in the direction of the minor axis. We did not deproject vertical distances for the inclination, so that the distance to the major axis as seen on the sky, $z$, is assumed to be the travelling distance of the cosmic rays. Since the galactic disc of IC~10 is tilted significantly away from edge-on, we potentially underestimate the travelling distances of the CREs by a factor of $1/\sin(i)=1.37$. However, simply deprojecting distances would overcorrect for the shortening because the projected width of the outflow $\cos(i)\times 1.4~{\rm kpc}=0.5~\rm kpc$ contributes to the observed height with an almost identical ratio: (halo height + projected outflow width) / (halo height) $=1.7~{\rm kpc}/1.2~{\rm kpc}=1.4$. This is corroborated by our synthetic intensity maps. The intensity scale height of the halo model in Fig.~\ref{fig:model} as measured on the sky is $0.45$~kpc; hence, the intrinsic scale height, taking into account line-of-sight integration of an inclined cylinder, corresponds to an effective correction for deprojection of $1.1$ which is close to $1.0$, so we simply use the distance as measured on the sky.

The profiles at 140 and 1580~MHz are fitted for an injection index, $\gamma_{\rm inj}$, where the CREs are injected in the midplane with a number density of $N(E,z=0)\propto E^{-\gamma_{\rm inj}}$. The CRE number density is evolved as function of distance from a solution of the diffusion--loss equation, where the losses are due to synchrotron and inverse Compton radiation as well as adiabatic losses \citep{heesen_18a}. Following this, the non-thermal radio continuum intensity is computed using the synchrotron emission spectrum of a single CRE. We assume a magnetic field strength of $B_0=12~\umu\rm G$ in the midplane, which comes from the assumption of energy equipartition between the cosmic rays and the magnetic field \citep{basu_17a}. Away from the midplane, we explore two basic magnetic field models: (i) we model the magnetic field strength everywhere else to fit the 140-MHz emission, so that as a consequence the 140-MHz intensities are fitted perfectly. A test for the quality of the model is then that the radio spectral index is fitted well. These magnetic field strengths are lower limits since our model assumes that the decrease of the intensities is mostly caused by the decrease of the magnetic field strength rather than CRE energy losses. (ii) we assume a magnetic field strength in approximate agreement with what is expected from energy equipartition. Finally, we assume a ratio of interstellar radiation field to magnetic energy density of $U_{\rm IRF}/U_{\rm B}=0.7$, which is the value found in the non-thermal superbubble \citep{heesen_15a}; this is also the global value if we use the hybrid H\,$\alpha + 24~\umu\rm m$ calibration \citep{kennicutt_12a} to calculate $\rm SFR_{\rm hyb}=0.052~M_{\sun}\,yr^{-1}$ and convert this to $U_{\rm IRF}$ with a galactocentric radius of $r_{\rm int}=0.7$~kpc \citep{heesen_16a}.
\begin{table}
\centering
\caption{Best-fitting cosmic-ray transport parameters for a CRE injection index of $\gamma_{\rm inj}=2.0$.\label{tab:parameters}}
\begin{tabular}{lcc}
\hline
 Parameter & North-eastern halo & South-western halo\\\hline
\multicolumn{3}{c}{Model~I: diffusion ($\umu=0.0$)}\\
$D_0$ & $2.2_{-0.6}^{+0.8}\times 10^{26}~\rm cm^2\,s^{-1}$ &  $5.0_{-1.2}^{+2.3}\times 10^{26}~\rm cm^2\,s^{-1}$\B\T\\
$\chi_{\rm red}^2$ & $5.5$ & $2.1$\B\\
dof & 10 & 13\T\\
\multicolumn{3}{c}{Model~I: diffusion ($\umu=0.5$)}\\
$D_0$ & $0.4_{-0.05}^{+0.05}\times 10^{26}~\rm cm^2\,s^{-1}$ &  $0.8_{-0.16}^{+0.15}\times 10^{26}~\rm cm^2\,s^{-1}$\B\T\\
$\chi_{\rm red}^2$ & $3.1$ & $1.5$\B\\
dof & 10 & 13\\
\multicolumn{3}{c}{Model~II: advection (constant wind speed)}\\
$V_0$ &  $20_{-3}^{+5}~\rm km\,s^{-1}$ & $29_{-5}^{+7}~\rm km\,s^{-1}$\B\T\\
$\chi_{\rm red}^2$ & $15$ & $6.3$\B\\
dof & 10 & 13\\
\multicolumn{3}{c}{Model~III: advection (accelerating wind)}\\
$V_0$ &  $14_{-3}^{+3}~\rm km\,s^{-1}$ & $23_{-4}^{+6}~\rm km\,s^{-1}$\B\T\\
$h_{V}$ & $0.6$~kpc & $0.8$~kpc\\
$\chi_{\rm red}^2$ & $16$ & $10$\B\\
dof & 10 & 13\\
\hline
\end{tabular}
\flushleft{{\bf Notes.}\\
$\umu$: parametrization of the energy dependence of the diffusion coefficient with $D=D_0(E/{\rm GeV})^{\umu}$.\\
$D_0$: diffusion coefficient at 1 GeV.\\
$\chi_{\rm red}^2$: reduced $\chi^2$ of the least squares-fitting with $\chi_{\rm red}=\chi^2/{\rm dof}$, where $\rm dof$ is the degrees of freedom.\\
$V_0$: advection velocity in the galactic midplane ($z=0$).\\
$h_{V}$: velocity scale height, so that the advection velocity is $V=V_0\times \exp(|z|/h_{\rm V})$.}
\end{table}

The minor axis profile of the radio spectral index between 140 and 1580 MHz shows the typical `parabolical' shape for diffusion \citep{heesen_16a}, so that diffusion models fit better than advection models. However, due to the mild inclination we cannot rule out that this is caused by a gradual transition from disc (young CRE) to halo (old CRE) emission. Hence, we have both fitted diffusion and advection models before we will return in Section~\ref{sec:discussion} to discuss both their advantages and issues. The CRE injection index is a free parameter and so are in addition the diffusion coefficient and the advection speed. Since we vary the magnetic field strength to fit the 140-MHz intensities, the number of free parameters is increased further.  We found that $\gamma_{\rm inj}=2.0$ provides the best fit to the data for all models since the non-thermal spectral indices between 140 and 1580~MHz are very flat with $\alpha_{\rm nt}\approx -0.5$. Increasing the injection index to $\gamma_{\rm inj}=2.1$ leads to poorer fits; as we cannot choose a lower injection index since then no spectral ageing occurs, we use $\gamma_{\rm inj}=2.0$ in what follows. For diffusion, we parametrized the diffusion coefficient as $D=D_0 (E /{\rm GeV})^{\umu}$ and tested models with either $\umu=0$ (no energy dependence) and with $\umu=0.5$, applicable for a Kolmogorov-type turbulence \citep{schlickeiser_02a}. In Table~\ref{tab:parameters}, we present our best-fitting cosmic-ray transport model parameters.

First, we found that for diffusion, energy dependent diffusion coefficients ($\umu=0.5$) fit slightly better than ones with no energy dependence ($\umu = 0$). We found that the magnetic field strength has only a small influence on the diffusion coefficients: for energy equipartition in the halo the diffusion coefficients increase by approximately 50 per cent. Second, we fitted advection models with a constant speed. Since advection results in approximately linear vertical spectral index profiles, the fit to the data is of lesser quality than for diffusion. For advection, we did not search for the formally best-fitting model, but for the one that best describes the spectral index decrease in the outskirts of the galaxy (the outermost one or two data points). The reasoning behind this choice is that these data points give us the cleanest estimate of the non-thermal radio spectral index, without the contamination from the emission near the galactic midplane. If we were to fit all data points instead, we would overestimate the advection speeds since the contribution to the radio continuum emission from younger CREs near the galactic midplane results in too flat non-thermal radio spectral indices. We found that the required advection speeds of $\approx$20--40~$\rm km\,s^{-1}$ are well below the escape velocity of $50~\rm km\,s^{-1}$, which we calculated with $V_{\rm esc} = \sqrt{2}\cdot V_{\rm rot}$, using a rotation speed of $V_{\rm rot}=36~\rm km\,s^{-1}$ \citep{oh_15a}. Again, the higher magnetic field strengths from energy equipartition do not change the results much: the advection speeds increase by approximately 50 per cent and are mostly still below the escape velocity.

Finally, we fitted advection models with an accelerating wind, where we assume $V(z)=V_0\times \exp(|z|/h_{V})$, which adds the velocity scale height, $h_{V}$, as an additional free parameter. Because the cosmic-ray number density decreases in an accelerating flow due to longitudinal stretching and adiabatic losses, the velocity scale height determines the scale height of the cosmic-ray energy density. We chose a velocity scale height of $0.6$--$0.8$~kpc, so that the cosmic rays are approximately in energy equipartition with the magnetic field and dominate over the magnetic field everywhere. We found again that the advection speed near the midplane is low, between 11 and 29~$\rm km\,s^{-1}$. But due to the acceleration, the escape velocity is exceeded at a height of $\approx$1~kpc above the midplane and at the edge of the observed halo, at $|z|=1.2$~kpc, the advection speed is $\approx$$100~\rm km\,s^{-1}$. For this model, we did not test the effect of energy equipartition magnetic fields since the field strength is close to energy partition anyway. The best-fitting non-thermal intensity profiles and spectral index profiles are shown in Fig.~\ref{fig:alpha}. 

\section{Discussion}
\label{sec:discussion}
Our best-fitting diffusion coefficients for $\umu=0.5$ are at least 2 orders of magnitude smaller than the values of a few $10^{28}~\rm cm^2\,s^{-1}$ that are found in the Milky Way and in other external galaxies for CREs with a few GeV, either from a spatial correlation analysis between the radio continuum emission and various star-formation tracers \citep[e.g.][]{berkhuijsen_13a} or from a spectral index analysis as in this work \citep{heesen_16a,mulcahy_16a}. The larger diffusion coefficients in spiral galaxies could be explained by their more ordered magnetic field structure, where the cosmic rays diffuse along magnetic field lines. This \emph{anisotropic} diffusion can possibly be suppressed in IC~10 due to its turbulent magnetic field structure. The ratio of the ordered to turbulent magnetic field strength, $q=B_{\rm ord}/B_{\rm turb}$, is as small as $0.06$ near the two giant \ion{H}{ii} regions \citep{chyzy_16a}. Cosmic-ray diffusion is effectively suppressed close to star-formation sites \citep{basu_17a}.

Away from the star-formation sites in the midplane, the field ordering is characterized by $q\approx 0.3$, comparable to what is found in the haloes of spiral galaxies, and anisotropic diffusion becomes a possibility. For a Kolmogorov-type turbulence, the diffusion coefficient parallel to magnetic field lines varies as $D_{\parallel}\propto q^{2} B_{\rm ord}^{-1/3} $ \citep{schlickeiser_02a}. In IC~10, $q=0.17$ \citep{chyzy_16a}, which is half of the value $q=0.3$ found in late-type spiral galaxies \citep{fletcher_10a}; also $B_{\rm ord} = 2~\umu\rm G$, which has to be compared with $B_{\rm ord}=10$--15~$\umu\rm G$ in the inter-arm region of spiral galaxies \citep{beck_16a}. Hence, we expect the \emph{anisotropic} diffusion coefficients to be an order of magnitude smaller than in spiral galaxies, or a few $10^{27}~\rm cm^2\,s^{-1}$, which is still at least an order of magnitude too high. As a consistency check, we can also calculate the diffusion coefficient using an estimate of the CRE diffusion length. The intensity scale height in the north-eastern halo ($z<-0.5~\rm kpc$) is $\approx$$0.35$~kpc and in the south-western halo ($z>0.5$~kpc) it is $\approx$$0.7$~kpc. Assuming energy equipartition and a non-thermal radio spectral index of $-0.5$, the CRE scale height is $h_{\rm e}=0.6$--$1.2$~kpc \citep[cf.\ equation~11 in][]{heesen_09a}. The CRE lifetime at 140~MHz is $t_{\rm rad}\approx40~\rm Myr$ near the midplane \citep[$B=12~\umu\rm G$, cf.\ equation~5 in][]{heesen_16a}. In the halo, the magnetic field strength is lower, so that the lifetime could be larger. However, it is unlikely that the age of the CREs exceeds the duration of the current star burst \citep[the age of the stellar clusters is 4--30~Myr;][]{hunter_01a}. Hence, using $D\approx h_{\rm e}^2/t_{\rm rad}$, we find values of $D\approx (3$--$6)\times 10^{27}~\rm cm^2\,s^{-1}$, which again are not in agreement with our values for energy dependence ($\umu=0.5$).
\begin{figure}
	\includegraphics[width=\hsize]{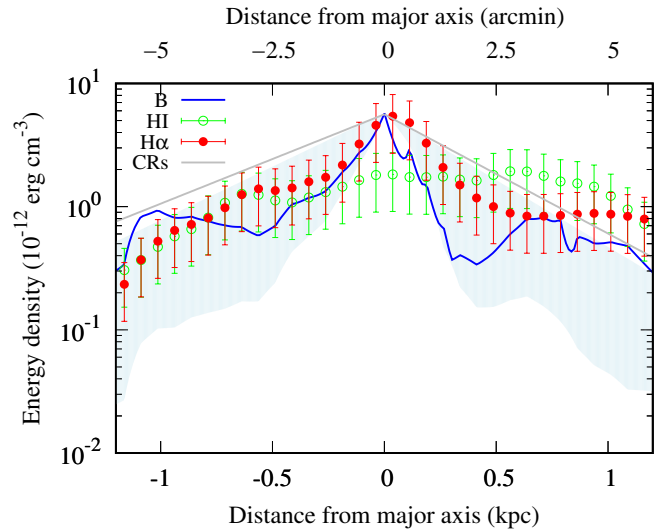}
    \caption{Minor axis profiles with distances from the major axis as seen on the sky. Shown are the energy densities of the magnetic field, warm neutral medium (\ion{H}{i}), warm ionized medium (H\,$\alpha$) and cosmic rays (CRs). The blue line shows the magnetic energy density for the accelerating wind model. The blue-shaded area indicates the range of magnetic field strengths for which we tested our models, where the upper limit is for energy equipartition and the lower limit is for almost freely escaping CREs (only radiation losses). Error bars show $\pm 1\sigma$ uncertainties, which are assumed to be 50 per cent stemming solely from the uncertainty of the velocity dispersion (see text for details).}
    \label{fig:density}
\end{figure}

Alternatively, we consider the possibility that the halo is advection dominated. Using a straightforward estimate for the advection speed of $V\approx h_{\rm e}/t_{\rm rad}$, we find speeds of $V\approx 15$--30$~\rm km\,s^{-1}$, in good agreement for our models with constant advection speeds. In this case, we find that the advection speeds would be well below the escape velocity of $\approx$$50~\rm km\,s^{-1}$. Hence, we have explored also the possibility of an accelerating wind. For this model, we have an additional free parameter in the velocity scale length. We found solutions, where the initial velocity in the midplane is similar to the velocity dispersion of both the warm ionized and warm neutral gas \citep[$20\pm 5~\rm km\,s^{-1}$;][]{thurow_05a,hunter_12a} and increases with height to exceed the escape velocity within 1~kpc distance from the midplane. The advantage of an accelerating wind is, besides the sufficient advection speeds, that the cosmic rays are in a self-consistent solution in approximate energy equipartition with the magnetic field. In our model, the cosmic-ray energy density is slightly larger than that of the magnetic field, which is expected if the cosmic rays are important in launching the wind. Increasing the cosmic-ray energy density in the halo would lead to slower accelerating winds; contrary, a lower cosmic-ray energy density would require that the wind accelerates faster. 

%\footnote{For the \ion{H}{i} emission the velocity dispersion is derived from the moment 2 map and for the H\,$\alpha$ emission from the line-width i.e.\ the variance of the emission}
Figure~\ref{fig:density} shows minor axis profiles of the averaged energy densities of the magnetic field, the warm ($\approx$$10^4$~K) neutral medium, traced by H\,{\sc i}, the warm ionized medium, traced by H\,$\alpha$, and of the cosmic rays. The magnetic energy density is $U_{\rm B}=B^2/(8\upi)$ and the kinetic energy density of the gas is $U_{\rm kin}=\rho \sigma_{\varv}^2/2$, where $\sigma_{\varv} = 20\pm 5~\rm km\, s^{-1}$ is the velocity dispersion. The gas density was measured from either the neutral or ionized hydrogen and includes a correction factor of $1.36$ in order to account for the contribution from Helium. Since we are interested in a comparison with our 1D cosmic-ray transport models, we neglect the error contribution from the variance of the density; thus, the error is solely determined by uncertainty of the velocity dispersion and is $\delta U_{\rm kin} = 2\delta \sigma_{\varv} /\sigma_{\varv}$, equivalent to 50 per cent. We find that the kinetic energy densities are within a factor of a few in energy equipartition with the magnetic field. This is the expected result since the magnetic field in dwarf galaxies is usually assumed to be amplified by the turbulent gas motions via the `fluctuating dynamo', which saturates when the energy densities become comparable \citep{schleicher_13a}. Interestingly in the central part, the magnetic energy density shows a peak that corresponds best to kinetic energy density of the ionized gas, whereas that of the atomic gas is practically constant. The cosmic-ray energy density is important everywhere. Such a behaviour has been suggested by theoretical works of cosmic-ray driven galactic winds \citep[e.g.][]{salem_14a}, where the cosmic rays can diffuse outwards from star-formation sites and thus create an overpressure in the surrounding area. A galactic wind can then be launched in the disc--halo interface, where the cosmic rays dominate the energy density and are able to transfer some of their energy and momentum to the ionized gas via the streaming instability.

While the diffusion model (Model~I) fits our data best, the resulting diffusion coefficients are at least one order of magnitude lower than both estimates for the expected diffusion coefficient. On the other hand, the projected extent of CRE injection sources due to the mild inclination can account for the parabolic shape of the spectral index profile, where the spectral index remains flat  out to distances of up to $0.5$~kpc away from the major axis; so the abrupt steepening of the spectral index at $|z|>0.5$~kpc that distinguishes diffusion from advection may be the result of geometry.  Hence, we explored the possibility of dominating advection and found that constant wind speeds are only a factor of two below the escape velocity -- the expected advection speed for a cosmic-ray driven wind \citep{breitschwerdt_93a}. The remaining discrepancy can be removed by the accelerating wind model (Model~III), which also provides us with a self-consistent solution where the magnetic energy density is similar to the kinetic energy densities of the warm neutral and warm ionized medium, while the cosmic rays are dominating by a factor of a few in the halo; this allows the cosmic rays to launch the wind. In summary, we favour Model~III that provides us with a physical self-consistent solution even though the fit to the data is poorer than for the diffusion model (Model~I).

\section{Conclusions}
\label{sec:conclusions}
In this paper, we have presented a low-frequency radio continuum study of the nearby dwarf irregular galaxy IC~10 using new LOFAR data. We have used the facet calibration technique in order to mitigate the effects of the Earth's ionosphere and aperture-array station beams on our observations. Thus, we obtained a deep 140-MHz radio continuum map, which we imaged at 44-arcsec resolution to bring out the weak, diffuse emission. We found that the emission far away from star-formation sites is not very prominent at these low frequencies because the non-thermal radio spectral index is flat, comparable to the injection spectrum. Nevertheless, we found that the radio continuum emission extends further along the minor axis than expected for the inclination angle of IC~10 and intensity profiles along the minor axis show a second, more extended, component away from the disc, which shows the expected behaviour for spectral ageing. We argue that this indicates the existence of a radio halo. The corresponding radio spectral index profile has a shape that can be best explained by diffusion. We then fitted the data with a stationary cosmic-ray diffusion model and found diffusion coefficients of $D=(0.4$--$0.8) \times 10^{26}(E/{\rm GeV})^{0.5}~{\rm cm^2\,s^{-1}}$. These small diffusion coefficients are likely to be found near star-formation sites, where the magnetic field is dominated by the turbulent component and the diffusion happens in an \emph{isotropic} way. In the halo, radio continuum polarimetry observations show a large-scale X-shaped field structure, which would allow the cosmic rays to diffuse \emph{anisotropically} along the magnetic field lines as well as be advected in a galactic wind \citep{chyzy_16a}. 

We hence have also explored cosmic-ray advection models, assuming that the typical parabolic shape of the radio spectral index profile, which is a telltale sign for diffusion, is the result of a gradual transition from disc to halo emission due to the mild inclination of the galaxy. We found that transport models with constant advection speeds result in best-fitting advection speeds well below the escape velocity. We have thus relaxed the assumption of a constant advection speed and found that our data can also be explained by an accelerating wind. In this model, the wind speed reaches the escape velocity at a height of less than $\approx$1~kpc above the midplane. The cosmic-ray energy density is everywhere important and the kinetic energy densities of the warm neutral and warm ionized medium are in approximate energy equipartition with the magnetic field. This fulfils the requirement of theoretical models for cosmic ray-driven winds, where the cosmic rays, the magnetic field and the ionized gas are advectively transported together. Our favourite model is hence that such a wind has developed in IC~10 over the past few tens of Myr -- the duration of the current star burst. This is corroborated by the advective time-scales at the edge of the halo ($|z|=1.2$~kpc) of 27~Myr in the south-western and 37~Myr in the north-eastern halo; they are in good agreement with 30~Myr, the age of the oldest stellar cluster \citep{hunter_01a}.

\section*{Acknowledgements}

We thank the anonymous referee for constructive comments that helped to improve the paper. LOFAR, the Low Frequency Array designed and constructed by ASTRON, has facilities in several countries, that are owned by various parties (each with their own funding sources), and that are collectively operated by the International LOFAR Telescope (ILT) foundation under a joint scientific policy. This research was performed in the framework of the DFG Forschergruppe 1254 Magnetisation of Interstellar and Intergalactic Media: The Prospects of Low-Frequency Radio Observations. FST acknowledges financial supports from the Spanish Ministry of Economy and Competitiveness (MINECO) under the grant number AYA2016-76219-P. AF thanks STFC (ST/N00900/1) and the Leverhulme Trust (RPG-2014-427).

%%%%%%%%%%%%%%%%%%%%%%%%%%%%%%%%%%%%%%%%%%%%%%%%%%



\bibliographystyle{mnras}
\bibliography{diff} % if your bibtex file is called example.bib

%%%%%%%%%%%%%%%%%%%%%%%%%%%%%%%%%%%%%%%%%%%%%%%%%%

% Don't change these lines
\bsp	% typesetting comment
\label{lastpage}
\end{document}